\documentclass[aps,pre,amsmath,amssymb,twocolumn,groupedaddress,showpacs]{revtex4}

\usepackage{graphicx}
\usepackage{amssymb}
\usepackage{amsmath}
\usepackage{epstopdf}
\newcommand{\be}{\begin{equation}}\newcommand{\ee}{\end{equation}}
\newcommand{\ba}{\begin{array}{l}}\newcommand{\ea}{\end{array}}
\newcommand{\baa}{\begin{eqnarray}}\newcommand{\eaa}{\end{eqnarray}}
\newcommand{\lab}[1]{\label{#1}}\newcommand{\re}[1]{(\ref{#1})}
\newcommand{\ci}[1]{\cite{#1}}

\begin{document}

\title{Quantum dynamics of PT-symmetrically kicked particle confined in a 1D box}
\author{J. Yusupov$^a$, S. Rakhmanov$^b$,  D. U.  Matrasulov$^a$ and H.
Susanto$^c$} \affiliation{
  $^a$ Turin Polytechnic University in Tashkent, 17
Niyazov Str., 100095,  Tashkent, Uzbekistan\\
$^b$National University of Uzbekistan, Vuzgorodok, Tashkent
100174,Uzbekistan\\
$^c$Department of Mathematical Sciences, University of Essex,
Wivenhoe Park, Colchester CO4 3SQ, UK}

\begin{abstract}
We study  quantum particle dynamics  in a box and driven by
PT-symmetric, delta-kicking complex potential. Such dynamical
characteristics as the average kinetic energy as function of time
and quasi-energy at different values of the kicking parameters.
Breaking of the PT-symmetry at certain values of the non-Hermitian
kicking parameter is shown. Experimental realization of the model
is also discussed.
\end{abstract}

\maketitle

\section{Introduction}

PT-symmetric quantum systems attracted much attention during past
two decades after the  discovery of the fact that non-Hermitian,
but PT-symmetric system can have a set of eigenstates with real
eigenvalues \ci{CMB1}.  In other words, self-adjointness of the
Hamiltonian is not necessary condition for being the eigenvalues
real. Currently quantum physics of PT-symmetric such systems has
become rapidly developing topic of contemporary physics  and great
progress is made in the study of different aspects of such
systems(see, e.g., papers \ci{CMB2}-\ci{ptbox} for review of
recent developments on the topic). These studies allowed to
construct complete theory of PT-symmetric quantum systems,
including PT-symmetric  field theory \ci{CMB07,CMB11}.
Experimental realization of such systems was also subject for
extensive research. The latter has been done mainly in optics
\ci{Makris,Nature,UFN,Konotop}. Some other PT-symmetric systems
are discussed recently in the literature \ci{Chit,Longhi}.
PT-symmetric relativistic system are also studied in
\ci{PTSD1,PTSD2}. General condition for PT-symmetry in quantum
systems has been derived in terms of so-called CPT-symmetric inner
product \ci{CMB5,CMB9,CMB11}. Similarly to the case of
Hermiticity, PT-symmetry in quantum systems can be introduced
either through the complex potential, or by imposing proper
boundary conditions, which provide such symmetry via the CPT-inner
product \ci{CMB5,CMB11}. Different types of complex potentials
providing PT-symmetry in Hamiltonian have been considered in
\ci{CMB9,CMB11,Kottos,Longhi}. PT-symmetric particle-in-box
system, where the box boundary conditions provide PT-symmetry of
the system, have been studied in \ci{CMB13,Znojil,Znojil2,ptbox}.
Certain progress is also done in nonlinear extension of
PT-symmetric systems \ci{UFN,Konotop,Panos}.

In this paper we consider quantum particle confined in a 1D box
and driven by a PT-symmetric, delta-kicking potential with the
focus on the role of non-Hermitian parameter on such
characteristics as everage kinetic, total energy and quasienergy.
Here we mention that some time ago, both the classical and the
quantum dynamics of systems interacting with a delta-kicking
potential have been extensively studied in the context of
nonlinear dynamics and quantum chaos theory
\ci{Casati}-\ci{Casati1}. Kicked quantum particle dynamics in a
box have been also considered in \ci{Roy,Well,Hu}. For kicked
systems, the classical dynamics is characterized by diffusive
growth of the average kinetic energy as a function of time, while
for corresponding quantum systems such growth suppressed (except
the special cases of so-called quantum resonances). The latter is
called quantum localization of classical chaos
\ci{Casati}-\ci{Casati1}. The dynamics of kicked nonrelativistic
system is governed by single parameter, product of the kicking
strength and kicking period.\\ We note that earlier,
PT-symmetrically kicked systems have been considered in the
Refs.\ci{Kottos,Longhi} in the context of quantum chaos theory. In
\ci{Kottos} PT-symmetrically kicked rotor is studied by developing
one-parameter scaling theory for non-Hermitian parameter and
focusing on the gain, loss effects. In \ci{Longhi}
PT-symmetrically kicked quantum rotor is studied by analyzing
quasienergy spectrum and evolution of the momentum distribution at
different values of the non-Hermitian parameter. Here we consider
PT-symmetrically kicked confined system, by focusing on the role
of confinement and non-Hermitian part of the kicking potential.
Usual way for creating of kicked quantum system is confining of
the system in a standing wave cavity. PT-symmetric analog of such
system could be realized in a cavity with the losses. Another
option,  putting the system in a transverse beam propagation
inside a passive optical resonator with combined phase and loss
gratings, was discussed, e.g., \ci{Longhi}. An optical waveguide
which is driven ny PT-symmetric optical field can be considered as
another version of the model we are going to treat. This paper is
organized as follows. In the next section we briefly recall
Hermitian counterpart of our system,  quantum particle confined in
a 1D box and driven by delta-kicking potential. In section III we
consider similar system with PT-symmetric delta-kicks. Section IV
presents some concluding remarks.
\begin{figure}[t!]
\includegraphics[totalheight=0.2\textheight]{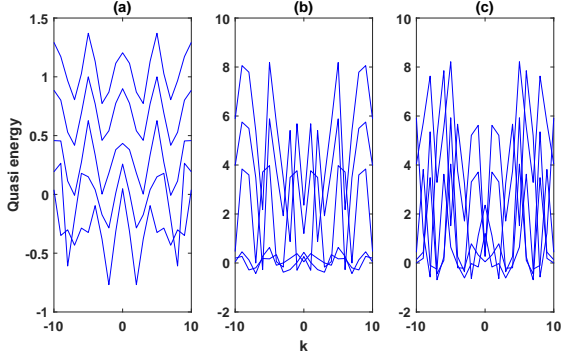}
 \caption{  Few quasienergy levels as a function of the wave number for different $K=\epsilon T$ , $K=0.1$ (a), $K=1$ (b),$K=1.5$ (c) and $L=10$ } \label{qe1}
\end{figure}

\section{Kicked quantum particle dynamics in a box}

Hermitian counterpart of the system we are going to study, is a
quantum particle confined in one-dimensional box of size $L$ and
driven by external delta-kicking potential given by
$$
U(x,t) =\epsilon \cos(\frac{2\pi x}{\mu}) \sum_l \delta(t-lT),
$$
where  $\mu,$ $\epsilon$ and $T$ are the wavelength, kicking
strength and period, respectively. Such system was considered
earlier in the context of quantum chaos theory e.g., in
\ci{Roy,Well,Li} and described by the following time-dependent
Schr\"{o}dinger equation:
\begin{equation}
i\frac{\partial}{\partial t} \Psi(x,t)
=\left[-\frac{1}{2}\frac{d^2}{dx^2} +U(x,t)\right]\Psi(x,t),
\label{kdb1}
\end{equation}
The wave function, $\Psi(x,t)$ fulfills the box boundary conditions
given by \be \Psi(0,t) = \Psi(L,t). \lab{bc01}\ee Exact solution of
Eq.\re{kdb1} can be obtained within the single kicking period
\cite{Casati,Well} by expanding the wave function, $\Psi(x,t)$ in
terms of the complete set of the eigenfunctions of the unperturbed
system as

\begin{equation}
\Psi(x,t) = \sum_n A_n(t) \psi_n(x) \label{1111}
\end{equation}
where $\psi_n(x)=\sqrt{2/L}\sin{(\pi n x/L)}$. Eqs.\re{kdb1} and
\re{1111}  lead to quantum mapping for the wave function amplitudes,
$A_n(t)$ which is given by
\begin{equation} A_n(t+T)=\sum_l A_l(t)
U_{ln}e^{-iE_l T}, \label{evol}
\end{equation}
where
$$
U_{ln} =\int_0^L \psi^*_n(x) e^{-i\epsilon \cos (2\pi x/\mu)}\psi_l
(x)dx.
$$
\begin{figure}[t!]
\includegraphics[totalheight=0.23\textheight]{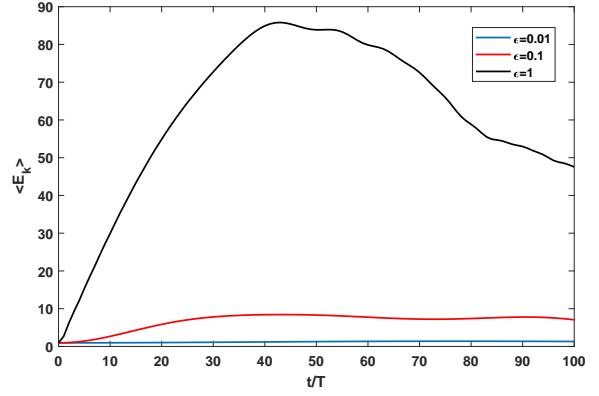}
 \caption{ (Color online) The average kinetic energy of kicked particle in a box as a function of kick number for different kicking strength for $L=3.3$, $T=0.01$ and $\mu=1.3$} \label{ke1}
\end{figure}
and \be E_l=(\pi l/L)^2 \lab{en1}\ee The amplitudes fulfill the norm
conservation given by \be N(t) = \sum_n |A_n(t)|^2 =1.
\lab{norm1}\ee We use this condition for controlling of the accuracy
of numerical computations. Thus the evolution of the wave function
within the single kicking period can be written as
$$
\Psi(x,t+T) = \hat U\Psi(x,t), \label{11111}
$$
where the one-period evolution operator is given by \be \hat U
=\exp(-i\frac{\partial^2}{2\partial x^2})\exp(-i\beta V(x))
\exp(-i\frac{\partial^2}{2\partial x^2}),\ee where
$$\beta =\frac{\pi T}{\mu^2}.$$
For such operator, one can consider the eigenvalue problem given
by \be \hat U\phi_n =\lambda_n\phi_n, \ee where the eigenvalues,
$\lambda_n$ are called quasienergy levels of the kicked system. In
Fig. \ref{qe1} few quasienergy levels are plotted as a function of
the wave number, $k =2\pi/\mu$ at different values of the
parameter $K=\epsilon T$. As $K$ is higher, as stronger the
fluctuations of the quasienergy levels.

Having found amplitudes and wave function, one can compute the
average kinetic energy, which is defined as
$$ <E(t)_{kin}> =
-\frac{1}{2}<\Psi(x,t)|\frac{d^2}{dx^2}|\Psi(x,t)>$$ \be=\sum_n
E_n|A_n(t)|^2, \lab{kinetic} \ee where $E_n$ are given by
Eq.\re{en1}. Fig. \ref{ke1} presents plots of the average kinetic
energy, $<E_k(t)>$ at different values of the kicking strength,
$\varepsilon$ for fixed kicking period $T$. Unlike the kicked rotor
$<E_k(t)>$ grows during some initial time and suppression with the
subsequent decrease occurs for large enough number of kick
($N=t/T$). For very large number of kicks one can observe periodic
or quasi-periodic time-dependence of $<E_k(t)>$. Such behavior in
some kicked quantum systems have been discussed in \ci{Hogg}.
Another feature of kicked quantum particle confined in a box is the
absence of quantum resonance. It should be noted that the dynamics
of kicked particle confined in a box depends on two factors, such as
interaction with the kicking force and bouncing of particle from the
box walls. Depending on the sign of of cosine in the kicking
potential, the kicking force can be attractive and repulsive. When
the kicking potential is repulsive particle gains the energy, while
in case of attractive potential it losses its energy. Therefore
depending on which area in the box, i.e. on the area where the
kicking force is positive or negative, acceleration or deceleration
of the particle may occur. Vey important factor is "synchronization"
of the kicking force and bouncing of particle from the box wall. It
also may cause acceleration and deceleration of the particle.

\begin{figure}[t!]
\includegraphics[totalheight=0.23\textheight]{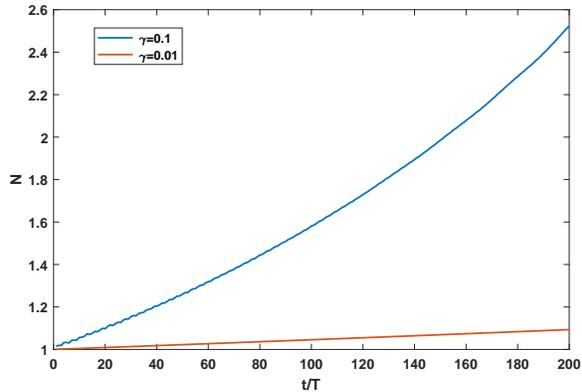}
 \caption{ (Color online) The norm as a function of kick number at different values of the $\gamma$ for $\epsilon=0.1$ ,  $L=3.3$ , $T=0.01$ and $\mu=1.3$} \label{norm01}
\end{figure}

\section{$\mathcal{PT}-$symmetrically kicked quantum particle  a one dimensional box}
PT-symmetric analog of the above system can be constructed by
adding into the kicking potential an imaginary part. Then
PT-symmetric kicking potential can be written as \be V_{PT}(x,t)
=f(t)\left[\epsilon \cos{(2\pi x/\mu)} + i \gamma \sin{(2\pi
x/\mu)}\right], \lab{ptk} \ee
 where $\epsilon$ and $T$ are the kicking strength and
period, respectively,  $\gamma\geq 0$ is the non-Hermitian
parameter that measures the strength of the imaginary part of the
potential and $ f(t)= \sum_l \delta(t-lT)$. The dynamics of the
system is governed by the following time-dependent Schr\"{o}dinger
equation:
\begin{equation}
i\frac{\partial}{\partial t} \Psi(x,t) =H_{PT}\Psi(x,t),
\label{perturbed1}
\end{equation}
where $H_{PT}$ is the Schr\"{o}dinger operator containing potential
$U_{PT}$. The same boundary conditions as those in Eq.\re{bc01}.
Exact solution of Eq.(\ref{perturbed1}) can be obtained similarly to
the case of Hermitian counterpart and one gets quantum mapping for
the evolution of the amplitude, $A_n(t)$ within the one kicking
period, $T$:
\begin{equation}
A_n(t+T)=\sum_l A_l(t) V_{ln}e^{-iE_l T}, \label{evol}
\end{equation}
where \be V_{ln} =\int \psi^*_n(x) e^{-i\epsilon \cos (2\pi
x/\mu)}e^{\gamma \sin (2\pi x/\mu)}\psi_l (x)dx \lab{kick2}\ee and
$E_l=(\pi l/L)^2$. The evolution operator corresponding to
Eq.\re{evol} can be written as \be \hat U_{PT}
=\exp(-i\frac{\partial^2}{2\partial x^2})\exp(-i\beta V(x))
\exp(-i\frac{\partial^2}{2\partial x^2}),\ee where
$$\beta =\frac{\pi T}{\mu^2}.$$

\begin{figure}[t!]
\includegraphics[totalheight=0.23\textheight]{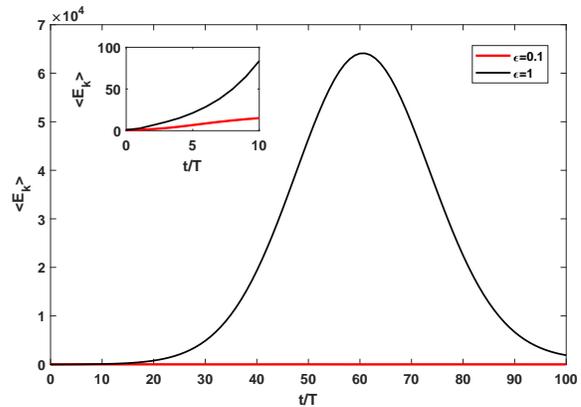}
 \caption{ (Color online) The average kinetic energy of $\mathcal{PT}-$symmetrically kicked particle in a box as a function of kick number for different kicking strength for $\gamma=0.1$, $L=3.3$ , $T=0.01$ and $\mu=1.3$} \label{ke2}
\end{figure}

\begin{figure}[t!]
\includegraphics[totalheight=0.2\textheight]{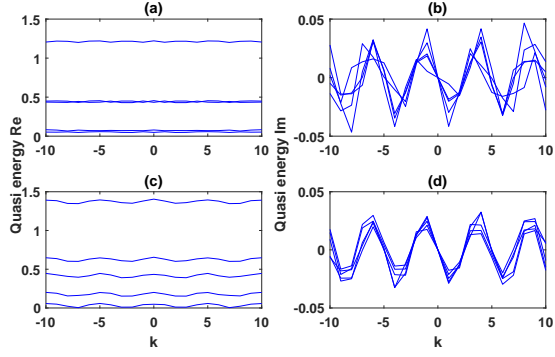}
 \caption{  The real part (a),(c) and imaginary (b),(d) parts of few quasienergy levels as a function of the wave number, $k=2\pi/\mu$, for $\gamma=1$ for $\epsilon=0.1$ (a),(b) and $\epsilon=1$ (c),(d) for $L=10$ } \label{qe2}
\end{figure}

\begin{figure}[t!]
\includegraphics[totalheight=0.2\textheight]{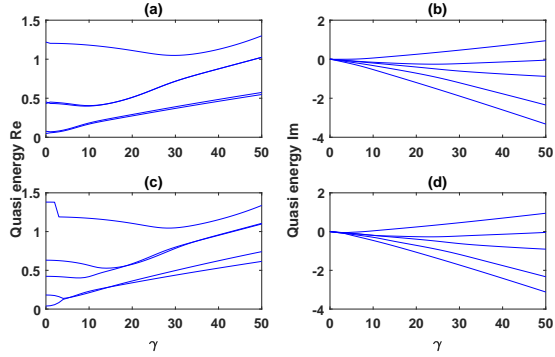}
 \caption{  The real part (a),(c) and imaginary (b),(d) parts few quasienergy levels as a function of
non-Hermitian parameter, $\gamma$ for $\epsilon=0.1$ (a),(b) and
$\epsilon=1$ (c),(d) for $L=10$}  \label{qe3}
\end{figure}

\begin{figure}[t!]
\includegraphics[totalheight=0.2\textheight]{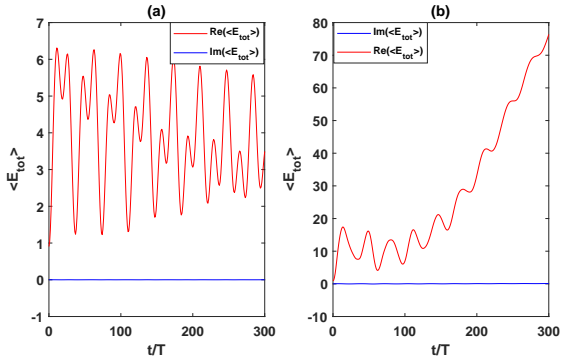}
 \caption{The average total energy computed as the expectation value of the operator $H_{PT}$ for $\gamma =0.01$ (a) and $\gamma =0.1$
 at $\epsilon =0.1$, $T=0.01$.}
 \label{tot1}
\end{figure}

For a quantum systems with complex PT-symmetric potentials, the norm
conservation is broken, i.e., the amplitudes, $A_n(t)$ do not
fulfill Eq.\re{norm1}.  Fig.\ref{norm01} presents plots of the norm
as a function of time at different values of $\gamma$ for fixed
$\epsilon$ and $T$. As higher is $\gamma$, as stronger is the
breaking of the norm conservation. In Fig. \ref{ke2} the average
kinetic energy is plotted as a function time. Although the profile
of plot is almost similar to that of Hermitian counterpart, the
values of $<E_k(t)>$ are much higher than that in Hermitian case.

Similarly to the Hermitian case, one can compute quasienergy
levels for PT-symmetric system as the eigenvalues of the operator
$U_{PT}$. Fig.\ref{qe2} presents few quasienergy levels as a
function of the wave number, $k=2\pi/\mu$. In Fig. \ref{qe3}  few
quasi-energy levels defined as the eigenvalues of the operator
$U_{PT}$  are plotted as a function of non-Hermitian parameter,
$\gamma$. Important feature of PT-symmetric systems with complex
potentials is the fact that is the expectation value of the
Hamiltonian operator is always real. This holds true also in case
of time-dependent operator. Fig. \ref{tot1} where the average
total energy, $<E_{tot}(t)>$ i.e. the expectation value of the
operator $H_{PT}$ is plotted as a function of time at different
values of $\gamma$.

\section{Conclusions}
We studied quantum dynamics of a particle confined in a 1d box and
driven by PT-symmetric, delta-kicking potential. Different
characteristics of the dynamics, such as the time-dependence of
the average kinetic energy, quasienergy and the average total
energy  are analyzed using the exact solution of the
time-dependent Schrodinger equation for single kicking period. It
is found that no unbound acceleration in PT-symmetric quantum
regime is possible, as the average kinetic energy is the periodic
or quasi-periodic in time. However,  in PT-symmetrically driven
system the gain of energy and acceleration are more intensive than
those for the Hermitian counterpart.   The above model can be
realized in different versions using optical systems where it is
possible to create PT-symmetric kicking potential. Such kicking
field could be realized e.g., in an  optical cavity with losses
and gains. Confining an optical pulse in such cavity would be a
version for our model. Another option is considering a
PT-symmetric periodic optical structure, e.g., array of optical
waveguides driven by laser field. In the absence of external
perturbation such system is described by the Helmholtz equation
with periodic boundary condition, which is an analog of the box
boundary condition. Therefore the driven waveguide array can be
considered as an analog of the above model.

\end{document}